\documentclass[]{article}

\pdfoutput=1

\usepackage{amsthm} 
\usepackage{bm} 
\usepackage{mathtools}
\usepackage{amsfonts}
\usepackage[authoryear]{natbib} 
\usepackage[letterpaper, margin=1in]{geometry} 
\usepackage{algorithm2e} 
\usepackage{float}
\allowdisplaybreaks 
\usepackage{graphicx} 
\usepackage{subcaption}
\usepackage{booktabs}
\usepackage{setspace}
\usepackage{authblk}

\newcommand{\Tra}{^{\sf T}} 
\newcommand{\Inv}{^{-1}} 
 
\newcommand{\V}[1]{{\bm{\mathbf{\MakeLowercase{#1}}}}} 
\newcommand{\Vtilde}[1]{{\bm{\tilde \mathbf{\MakeLowercase{#1}}}}} 
\newcommand{\Vhat}[1]{{\bm{\hat \mathbf{\MakeLowercase{#1}}}}} 
\newcommand{\M}[1]{{\bm{\mathbf{\MakeUppercase{#1}}}}} 
\newcommand{\Mtilde}[1]{{\bm{\tilde \mathbf{\MakeUppercase{#1}}}}} 

\def\A{{\bf A}} 
\def\a{{\bf a}} 

\def\B{{\bf B}} 
\def\Btilde{\Mtilde{B}}
\def\b{{\bf b}} 
\def\btilde{\Vtilde{b}}

\def\C{{\bf C}} 
\def\c{{\bf c}}

\def\E{{\bf E}} 
\def\e{{\bf e}} 
\def\ehat{{\Vhat{e}}}

\def\G{{\bf G}}
\def\g{{\bf g}}

\def\I{{\bf I}}

\def\R{{\bf R}}

\def\W{{\bf W}} 
 
\def\Wtilde{\Mtilde{W}}

\def\X{{\bf X}} 
 
\def\Xtilde{\Mtilde{X}}

\def\Y{{\bf Y}}
\def\y{{\bf y}}

\def \valpha{{\V{\alpha}}}
\def \vbeta{{\V{\beta}}}
\def \vpi{{\V{\pi}}}

\def \mlambda{{\M{\Lambda}}}

\def \mtheta{{\M{\Theta}}} 
\def \vtheta{{\V{\theta}}}

\def \msigma{{\M{\Sigma}}}

\newcommand{\normal}{\mathcal{N}}

\newcommand{\simiid}{\overset{iid}{\sim}}

\title{Simultaneous Variable Selection, Clustering, and Smoothing in Function on Scalar Regression} 
\author[1]{Suchit Mehrotra}
\author[1]{Arnab Maity}
\affil[1]{Department of Statistics, North Carolina State University}

\begin{document}

\maketitle

\begin{abstract} 
    We address the problem of multicollinearity in a function-on-scalar regression model by
    using a prior which
    simultaneously selects, clusters, and smooths functional effects. Our methodology groups
    effects of highly correlated predictors, performing dimension reduction
    without dropping relevant predictors from the model. We validate our
    approach via a simulation study, showing superior performance
    relative to existing dimension reduction approaches in the function-on-scalar
    literature. We also demonstrate the use of our model on a data set
    of age specific fertility rates from the United Nations Gender
    Information database. 
\end{abstract}

\section{Introduction}

\onehalfspacing

Recent technological advances in computing power and data storage have
simplified the collection of vast quantities of data. Modern scientific
questions often involve statistical analyses on datasets with a large number of
predictors, many of which have no meaningful effect on the response. In numerous
situations, an analysis is further complicated by multicollinearity in
predictors. An example are microarray studies where interest lies in
understanding the relationship between a health outcome and gene expressions;
gene expression levels are often highly correlated and the number of genes is
substantially larger than the number of samples. In such situations, modern statistical approaches have 
utilized the assumption that highly
correlated predictors share the same effect on the response, especially if the correlation
is due to the same underlying factor. They aim to reduce the dimension of the
problem by summing or averaging groups of highly correlated columns and fitting
a linear model with the new set of predictors, which, in essence, is equivalent to finding clusters of regression coefficients in a supervised manner. 

Many methods exist which aim to cluster coefficients in the univariate linear model.
For example, \cite{bondell2008oscar}, \cite{she2010sparse}, and
\cite{sharma2013pacs} all propose methods which minimize a loss function plus a
penalty term to encourage clustering in the coefficients. These methods are
computationally efficient as they are all a special case of the generalized
lasso \citep{tibshirani2011genlasso}. In the Bayesian setting, the coefficient
clustering problem is solved by utilizing mixture models as priors for the
coefficients. \cite{tadesse2005bayesian} uses finite mixture models to cluster
the coefficients, while \cite{nott2008predictive}, \cite{kim2006variable},
\cite{dunson2008bayesian}, and \cite{curtis2011bayesian} all use Dirichlet
Process priors to estimate clusters. 

In this paper we add to the existing dimension reduction literature for models
with a functional response and a set of scalar predictors, which up till now has
focused only on variable selection. At present, many group penalties exits to
select variables in a function-on-scalar regression. First, the group lasso
\citep{yuan2006group} can be used to shrink basis function
coefficients belonging to the same predictor simultaneously to zero.
\cite{wang2007group} modify this approach by using a group SCAD penalty,
\cite{chen2016variable} use a group MCP penalty, \cite{fan2017high}
propose an adaptive group lasso, while \cite{kowal2018bayesian} utilize a group horseshoe prior. 
However, an issue with these approaches is the
need to fix the number of basis functions before fitting the model, and recent work has focused on
smoothing the coefficient functions along with selecting variables.  For
example, \cite{goldsmith2016assessing} use a penalized normal prior to
smooth coefficient functions without any variable selection, while
\cite{parodi2018} propose a framework for simultaneous variable selection and
smoothing. 

Currently, to the best of our knowledge, no function-on-scalar approaches exist which
address multicollinearity in the predictors by exploiting a cluster structure in
the coefficient functions. Additionally, in the Bayesian setting, methods to
simultaneously select and smooth coefficient functions have yet to be developed. Consequently, our manuscript contributes to the current
function-on-scalar literature by proposing a novel clustering based dimension reduction
technique for functional data, heretofore unexplored. 

Our paper proceeds as follows. We review the relevant background literature in
Section \ref{sec:background}, our model and relevant computational details in
Sections \ref{sec:model} and \ref{sec:computation}, a simulation study and its
results in Section \ref{sec:simulation}, along with a real data application in
Section \ref{sec:application}. 

\section{Background} \label{sec:background}

In a univariate linear model, an approach to finding clusters in a vector of
coefficients, $\vbeta$, is to assume the existence of $K$ clusters and use
finite mixture models as a prior for $\vbeta$ \citep{tadesse2005bayesian}. Using
a latent class representation for cluster membership the prior can be written
as:
\begin{align} \label{eqn:fmm_reg}
\begin{split}
(\forall p \in \{1, \dots, P \}) \ \beta_p | c_p, \vtheta & =
\sum_{k = 1}^K I(c_p = k) \theta_k \\
(\forall p \in \{1, \dots, P \}) \ c_p | \vpi & \sim 
\text{ Discrete}(\pi_1, \dots, \pi_K) \\
(\forall k \in \{1, \dots, K \}) \ \theta_k | \lambda & \sim \normal(0, \lambda^{-1})
\end{split}
\end{align}
where $\c$ is a $P \times 1$ vector which stores class labels for the elements of
$\vbeta$. The hierarchy in \eqref{eqn:fmm_reg} implies that $\beta_p \sim G =
\sum_{k = 1}^K \pi_k \delta_{\theta_k}(\beta_p)$ where $\delta_{\theta_k}(x)$ is
1 if $x = \theta_k$ and $0$ otherwise. Consequently, each element of $\vbeta$ is
a finite mixture of K delta functions, and can only take on one of $K$ values.
If we know $K$ in advance, estimating $\vtheta$ and class memberships, $\c$,
provides a solution to the coefficient clustering problem. 

However, this approach suffers from a major drawback in that $K$ has to be fixed
a-priori, which is information rarely available in applied settings. To remedy
this issue, one can use priors which allow for a countably infinite number of
mixture components, such as the Dirichlet Process. 

\subsection{Dirichlet Process Priors} 

The Dirichlet Process (DP) is indexed by a concentration parameter $\alpha$ and
a base distribution $G_0$, denoted $DP(G_0, \alpha)$.  It has a long history of
being used as a clustering prior for a set of points, $\{x_1, \dots x_N\}$, and
was first investigated by \cite{ferguson1973} and \cite{antoniak1974}. More
recently, computationally efficient methods have been developed for parameter
estimation by \cite{escobar1995bayesian}, \cite{maceachern1998estimating}, and
\cite{neal2000markov}. Using a stick breaking construction of the Dirichlet
process, it can be shown that  $G(x) = \sum_{k = 1}^\infty \pi_k
\delta_{\theta_k}(x)$ is a sample from a $DP(G_0, \alpha)$, where $\pi_k = V_k
\prod_{l = 1}^{k - 1} (1 - V_l)$, $V_k \simiid \text{Beta}(1, \alpha)$, and
$\theta_k \simiid G_0$ \citep{sethuraman1994constructive}. Hence, the
distributions sampled from a Dirichlet Process are a countably infinite mixture
of point masses. 

Key insights into the behaviour of the Dirichlet Process were given by
\citet{blackwell1973ferguson} and \cite{neal2000markov}.  \cite{neal2000markov}
derives a conditional prior for the class labels, $\c$, by viewing the Dirichlet
Process mixture model as the limit of finite mixture models as $K \to \infty$.
By integrating out the mixing parameters $\vpi$ he gives the following
conditional priors for the elements of $\c$
\begin{align}
\label{eqn:neal_prior}
\begin{split}
	P(c_i = k | \c_{-i} ) = \frac{n_{-i, k}}{N - 1 + \alpha} \\
	P(c_i \neq c_j \text{ for all } i \neq j | c_{-i}) = 
	\frac{\alpha}{N - 1 + \alpha}
\end{split}
\end{align}
where $n_{-i, k}$ is the number of $c_j$, $j \neq i$, that are equal to $k$ and $\c_{-i}$
is $\c$ without the $i^{th}$ element.
This representation allows us to bypass the sampling of mixing weights, $\vpi$,
and shows that the prior conditional probability of an observation being
assigned to an existing cluster is proportional to the number of elements in that
cluster. 

For the univariate linear model, we can modify the hierarchy in
\eqref{eqn:fmm_reg} for the Dirichlet Process prior as follows:
\begin{align}
\label{eqn:lm_dp}
\begin{split}
	(\forall p \in \{1, \dots, P \}) \ \beta_p | G & \sim G \\
	G & \sim DP(G_0, \alpha) \\
	G_0 | \lambda & = \normal(0, \lambda\Inv) 
\end{split}
\end{align}
\citet{nott2008predictive} explored the estimation and predictive capacity of
the DP prior with a normal base measure along with its applications in penalized
spline smoothing. He also derived a Gibbs sampler for the model in
\eqref{eqn:lm_dp} by leveraging a latent class representation for group
membership. His algorithm iterates between sampling the cluster indicators for
the elements of $\vbeta$, $\c$, and all other model parameters conditioned on
the number of clusters in $\c$, $K$. We leverage this algorithm for development
of our own Gibbs samplers and will discuss its relevant details in Section
\ref{sec:computation}.

\subsection{Variable Selection and Clustering}

\cite{kim2006variable}, \cite{dunson2008bayesian}, and \cite{curtis2011bayesian}
propose extensions to the prior in \eqref{eqn:lm_dp} for simultaneous variable
selection and clustering.  They use a mixture of a point mass at zero with a
Dirichlet process prior \begin{align} \label{eqn:vs_clust} G = \pi_0 \delta_0 +
(1 - \pi_0) G^*, \ G^* \sim DP(G_0, \alpha) \end{align} where $0 \leq \pi_0 \leq
1$ is a mixture weight, $\delta_0$ is a point mass probability density at $0$,
and $DP(G_0, \alpha)$ is a Dirichlet Process prior with precision $\alpha$ and
base measure $G_0$. Using the constructive definition of
\citet{sethuraman1994constructive} the prior in \eqref{eqn:vs_clust} is again an
infinite mixture of point masses with the first point mass fixed at zero
\begin{align} \label{eqn:vs_clust_inf}
    G = \pi_0 \delta_0 + (1 - \pi_0) \sum_{k = 1}^{\infty} \pi_k
        \delta_{\theta_k} = \sum_{k = 0}^{\infty} \tilde{\pi}_k \delta_{\theta_k}
\end{align}
where $\tilde{\pi}_0 = \pi_0$, $\tilde{\pi}_k = (1 - \pi_0) \pi_k$ if $k \geq
1$, $\theta_0 = 0$ and $\theta_k \simiid G_0$ if $k \geq 1$.  Computation for
this model is similar to the model discussed in \eqref{eqn:lm_dp} because the
prior in \eqref{eqn:vs_clust_inf} can be interpreted as a clustering prior with
the restriction that the first cluster is zero. We can account for this
restriction by making a few changes to the Gibbs sampler developed by
\citet{nott2008predictive}. 

\section{Model} \label{sec:model}

In this section we extend the methods discussed in Section \ref{sec:background}
to models with a functional response and a set of 
scalar covariates. Let $i \in \{1, \dots, N\}$ be the subject
level index, and $p \in \{1, \dots, P \}$ be the predictor index. Then the general
function-on-scalar model is: 
\begin{align} \label{eqn:fosr} 
	y_i(t) = \mu(t) + \sum_{p = 1}^P x_{ip} \beta_p(t) + e_i(t)
\end{align}
where $y_i(t)$ is the functional response value for subject $i$ at time $t$,
$\mu(t)$ is an intercept term, $x_{ip}$ is the $p^{th}$ predictor value for the
$i^{th}$ individual, $\beta_p(t)$ is the functional effect for the $p^{th}$
predictor at time $t$, and $e_i(t)$ is an error term. If all observations are
observed on a common grid, this model can be written in matrix form as: $\Y = \X
\vbeta\Tra + \E$. 

Since we wish to cluster only some of the predictors (at the least the intercept
will remain unclustered), we delineate two groups of functional effects and
propose the following model for a grid of equally spaced points: 
\begin{align} \label{eqn:dense_model}
    \Y = \W \valpha\Tra + \X \vbeta\Tra + \E
\end{align} 
where $\Y_{N \times T}$ is a matrix of responses, $\W_{N \times P_f}$ is a
matrix of predictors with free effects, $\valpha_{T \times P_f}$, $\X_{N \times
P_c}$ is a matrix of predictors whose effects, $\vbeta_{T \times P_c}$, we wish
to select and cluster, and $\E_{N \times T}$ is a matrix of errors with the
i$^{th}$ row $\e_i \sim \normal_T(\V{0}, \tau\Inv \I)$. We expand each
coefficient function as a linear combination of $M$ b-spline basis functions,
$\{\theta_1, \dots, \theta_M \}$, such that $\valpha_{T \times P_f} = \mtheta_{T
\times M} \A_{M \times P_f}$ and $\vbeta_{T \times P_c} = \mtheta_{T \times M}
\Btilde_{M \times P_c}$. This reduces the functional estimation problem to
estimating the elements of $\A$ and $\Btilde$. 

In practice, care must be taken in choosing $M$, balancing the need for a
function to remain flexible with the risk of overfitting. We address this
issue by expanding the functional effects using more basis functions than we
think are necessary and use penalization to induce smoothness in the coefficient
function. We follow \cite{goldsmith2016assessing} and use a full-rank penalty
matrix $\R$ to penalize basis coefficients, setting $\R = \eta \I + (1 - \eta)
\R_2$, where $\I$ is the identity matrix, $\R_2$ is a second degree P-spline
penalty \citep{eilers1996flexible}, and 
$\eta = 0.001$. Consequently, the prior for the columns of $\A$, $\a_p$, is
$\forall p \in \{1, \dots, P_f \}$
\begin{align} \label{eqn:smooth_prior}
	 \a_p & \sim \normal_{M}(\V{0}, \lambda_{a_p}\Inv \R\Inv )
\end{align}
where $\lambda_{a_p} \sim \mathcal{G}(a_\lambda, b_\lambda)$ and $\mathcal{G}(a,
b)$ is the Gamma distribution with shape $a$ and rate $b$. 

\subsection{Clustering Functional Effects}

Many approaches exist for clustering functional data and we refer the reader to
\cite{jacques2014functional} for a review. Because we expand each coefficient
function using b-splines, our clustering approach is similar to the work
initially proposed by \cite{abraham2003unsupervised}; we find clusters in
$\vbeta$ by clustering the coefficients of the basis expansion. If we assume the
columns of $\vbeta$ come from $K$ clusters, we can expand it as $\vbeta =
\mtheta \Btilde = \mtheta \B \C\Tra$ where $\mtheta_{T \times M}$ and $\B_{M
\times K}$ are the basis function and coefficient matrices, respectively, and
$\C_{P_c \times K}$ is a matrix whose rows are one-hot encoded with class
membership information.  Consequently, $\X \vbeta\Tra = \X \Btilde\Tra
\mtheta\Tra = \X \C \B\Tra \mtheta\Tra$, and $\C$ has the effect of summing
columns of $\X$ with the same effect on the response. 

Because we are also interested in estimating smooth coefficient functions, we
give each predictor its own smoothing parameter, $\tilde{\lambda}_{b_p}$.
Therefore, our task is to estimate a mean and smoothing parameter for each of
our clusters. We can easily do this by modifying our base distribution to be a
multivariate normal-gamma \citep{ray2006functional, zhang2014joint}. Consequently, for clustering functional effects, we
propose the prior:
\begin{align} \label{eqn:prior_dp}
\begin{split}
	[\forall \in \{1, \dots, P_c \}] \ \ (\btilde_p, \tilde{\lambda}_{b_p}) & \sim G \\
	G & \sim DP(G_0, \alpha) \\
	G_0 & = \normal_{M}(\V{0}, \tilde{\lambda}_{b_p}\Inv \R\Inv) \mathcal{G}(a_{\lambda}, b_{\lambda})
\end{split}
\end{align}
Extending this prior for simultaneous variable selection and clustering requires
only one modification: 
\begin{align} \label{eqn:prior_dppm}
\begin{split}
	[\forall \in \{1, \dots, P_c \}] \ \ 
        (\btilde_p, \tilde{\lambda}_{b_p}) & \sim 
        \pi_0 \delta_{\V{0}} + (1 - \pi_0) G^* \\
	G^* & \sim DP(G_0, \alpha) \\
	G_0 & = \normal_{M}(\V{0}, \tilde{\lambda}_{b_p}\Inv \R\Inv) 
        \mathcal{G}(a_{\lambda}, b_{\lambda}) \\
	[\pi_0, (1 - \pi_0)] & \sim \text{Dirichlet} 
        \left( \frac{\alpha_0}{2}, \frac{\alpha_0}{2} \right) 
\end{split}
\end{align}
In all computations in this paper we give $\alpha$ a non-informative Gamma prior
and update it using the method outlined in \cite{escobar1995bayesian}. We also
fix $\alpha_0 = 2$ so that $\pi_0 \sim \text{Unif}(0, 1)$. 

\section{Computation} \label{sec:computation}

In this section we give the details of a Gibbs sampler for estimating the
parameters of our proposed models. The crux of the sampler is the update of the
class labels for each parameter, $\c$, and we extend the approach used by
\cite{nott2008predictive} to the model in \eqref{eqn:dense_model}. 

We first list the model hierarchy conditioned on the cluster indicator
matrix, $\C$, and assume that it has $K$ columns.  We also modify
\eqref{eqn:dense_model} and work with the transposed equation, $\Y\Tra = \mtheta
\A \W\Tra + \mtheta \B (\X \C)\Tra + \E\Tra$. This, combined with the identity
$\text{vec}(\A \B \C\Tra) = (\C \otimes \A) \text{vec}(\B)$, implies that our
model can be written as $\y = \Wtilde \a + \Xtilde \b + \e$ where $\y =
\text{vec}(\Y\Tra)$, $\e = \text{vec}(\E\Tra)$, $\a = \text{vec}(\A)$, $\b =
\text{vec}(\B)$, $\Xtilde = (\X \C) \otimes \mtheta$ and $\Wtilde = \W \otimes
\mtheta$. Therefore, conditioned on $\C$, our model hierarchy is given by: 
\begin{align} \label{eqn:cond_hier}
\begin{split}
	\y & \sim \normal_{NT}(\Wtilde \a + \Xtilde \b, \tau\Inv \I) \\
	\a & \sim \normal_{MP_f}(\V{0}, \mlambda_a\Inv \otimes \R\Inv) \\
	\b & \sim \normal_{MK}(\V{0}, \mlambda_b\Inv \otimes \R\Inv) \\
	[\forall p \in \{1, \dots, P_c \}] \ \ \lambda_{a_p} & \sim 
        \mathcal{G}(a_{\lambda}, b_{\lambda}) \\
	[\forall p \in \{1, \dots, K \}] \ \ \lambda_{b_p} & \sim 
        \mathcal{G}(a_{\lambda}, b_{\lambda}) \\
	\tau & \sim \mathcal{G}(a_{\tau}, b_{\tau})
\end{split}
\end{align}
where $\mlambda_a = \text{diag}(\lambda_{a_1}, \dots, \lambda_{a_{P_f}})$ and
$\mlambda_b = \text{diag}(\lambda_{b_1}, \dots, \lambda_{b_K})$. From \eqref{eqn:cond_hier} it is easy to see that the updates of the
parameters are similar to those for a normal linear model, with the posteriors
of $\a$ and $\b$ being multivariate normal, and the posteriors of the
$\lambda$'s and $\tau$ being Gamma. Therefore, the rest of this section will
focus on estimation of the cluster indicator matrix, first for the clustering
only Dirichlet Process prior \eqref{eqn:prior_dp} followed the variable
selection and clustering prior \eqref{eqn:prior_dppm}. 

\subsection{Dirichlet Process} \label{sec:dp}

Updating $\c$ is the most computationally intensive step of the Gibbs
sampler because its elements have to be updated sequentially.  To update the
class label for a predictor, $c_p$, we need to calculate posterior probabilities
of the predictor belonging to each cluster. In the linear model setting,
\citet{nott2008predictive} updates $c_p$ by integrating out the parameter
vector, which is important for the mixing of the Markov chains, but requires $K
+ 1$ matrix inversions for each predictor. Here, $K$ is the number of unique
elements in $\c_{-p}$ and the additional inversion is for the proposal of a new
cluster. Hence, his algorithm is a modification of the collapsed Gibbs sampler; 
Algorithm 3 in \cite{neal2000markov}. 

For our purposes, the inclusion of a prior on the smoothing parameters in
\eqref{eqn:prior_dp} complicates the integral required to collapse the over all
cluster parameters. Hence, we combine Algorithms 3 and 8 from
\cite{neal2000markov} and collapse only over the cluster means. Our update of
the elements of $\c$ proceeds as follows. For each $c_i$, $i \in \{1, \dots, P_c \}$:
\begin{itemize}
	\item Let $k^{-}$ be the number of distinct $c_j$ for $j \neq i$ and 
	$h = k^{-} + 1$.
	\item If $c_i = c_j$ for some $j \neq i$ draw from 
	$\mathcal{G}(a_{\lambda}, b_{\lambda})$ for $\lambda_{b_h}$. 
	If $\forall j \neq i$, $c_i \neq c_j$, set $\lambda_{b_h} = \tilde{\lambda}_{b_i}$, where $\tilde{\lambda}_{b_i}$ is 
	the 
	smoothing
	parameter for the $i^{th}$ predictor. 
	\item For each $k \in \{1, \dots, h\}$
	\begin{align*}
		P(c_p = k | \c_{-p}, \V{\lambda}_b, \cdot) & 
		\propto f(\y | \c', \V{\lambda}_b, \cdot) 
		P(c_p = k | \c_{-p}, \alpha) \\
		& = \int f(\y | \b, \c', \V{\lambda}_b, \X, \W, \a, \tau) 
			f(\b | \V{\lambda}_b, \R) \ d \b  
			\times P(c_p = k | \c_{-p}, \alpha)
	\end{align*}
\end{itemize}
where 
\begin{align} \label{eqn:c_int}
\begin{split}
& \int f(\y | \b, \c', \V{\lambda}_b, \X, \W, \a, \tau) 
	f(\b | \V{\lambda}_b, \R) \ d \b = \\
	& \ \ \ \ \ \ \ \ \ \ \ \ \ \ \ \ \ \ \ \ 
	(2 \pi)^{- \frac{NT}{2}} \tau^{\frac{NT - M\tilde{K}}{2}}
|\mlambda_b'|^{\frac{M}{2}} |\R|^{\frac{\tilde{K}}{2}} |\G|^{-\frac{1}{2}} 
\exp \left\{-\frac{\tau}{2} ( \ehat_W\Tra \ehat_W - \g\Tra\G\Inv\g) \right\}
\end{split}
\end{align}
and $\G = [(\X \C') \otimes \mtheta]\Tra [(\X \C') \otimes \mtheta] +
\frac{1}{\tau} (\mlambda_b' \otimes \R)$, $\g = [(\X \C') \otimes \mtheta]\Tra
\ehat_W$, $\ehat_W = \text{vec}(\Y\Tra - \mtheta \A \W\Tra)$, 
$\c'$ is the vector of class labels with 
$c_p = k$, $\C'$ is
the corresponding one-hot encoded matrix, $\tilde{K}$ is the number of clusters
in $\c'$, $\b$ is a $M \tilde{K} \times 1$ vector, $\mlambda_b'$ is a
$\tilde{K} \times \tilde{K}$ dimensional diagonal matrix of smoothing parameters
for clusters present in $\c'$, and $P(c_p = k | \c_{-i}, \alpha)$ can be
calculated by using the formulas in \eqref{eqn:neal_prior}. It should be noted
that while $h$ inverses have to be computed for each predictor, they can be done
in parallel to speed up the algorithm.
 
\subsection{Dirichlet Process + Point Mass} \label{sec:dppm}

Computation for this model is similar to the model discussed in Section
\ref{sec:dp} because the prior in \eqref{eqn:prior_dp} is a
clustering prior with the restriction that the first cluster is zero. We can
account for this by making a few changes to the Gibbs sampler
outlined above. To update $\c$, conditional prior
probabilities of the form in \eqref{eqn:neal_prior} need to be determined and
the integral in \eqref{eqn:c_int} has to be modified by dropping, from $\X$, the
columns in the null (zero) cluster. The update of $\Btilde$ also relies on
dropping the columns in the null cluster because their effects are restricted
to be zero; the rest of the update proceeds as before with the smaller
predictor matrix. 

If we let $P_0$ denote the number of variables which are in the null cluster and
$P_{nz}$ denote the number of variables which are non-zero $(P_c = P_0 + P_{nz})$, we
can calculate conditional prior probabilities of the form in
\eqref{eqn:neal_prior} by thinking of the prior in two levels: the first being a
two component finite mixture model and the second a Dirichlet Process. To begin,
we need to determine the prior probabilities of the predictor $p$ being equal to
zero, $P(c_p = 0 | \c_{-p}, \alpha_0)$, where $[\pi_0, (1 - \pi_0)]$ is given a
$ \text{Dirichlet} \left( \frac{\alpha_0}{2}, \frac{\alpha_0}{2} \right)$ prior.
This is given directly by \citet{neal2000markov} to be 
\begin{align} \label{eqn:fm_dp}
    \pi^*_0 = \pi(c_p = 0 | \c_{-p}, \alpha_0) & = 
    \frac{n_{-p, 0} + \alpha_0 / 2}{P_c - 1 + \alpha_0}
\end{align}
Because only a subset of the predictors are non zero, the Dirichlet Process
prior can be thought of as active for only $P_{nz}$ predictors. Therefore, the
probabilities from \eqref{eqn:fm_dp} can be combined with the probabilities from
\eqref{eqn:neal_prior} to get the conditional prior for cluster membership:
\begin{align}
\begin{split}
    \pi^*_0 = P(c_p = 0 | \c_{-p}, \alpha_0) & = 
        \frac{n_{-p, 0} + \alpha_0 / 2}{P_c - 1 + \alpha_0} \\
    \text {For $k \geq 1$, } P(c_p = k | \c_{-p}, \alpha) & = 
        (1 - \pi_0^*) \left( \frac{n_{-p, k}}{P_{nz} - 1 + \alpha} \right) \\
    P(c_p \neq 0 \text{ and } 
        c_p \neq c_j \text{ for all } p \neq j| \c_{-p}, \alpha) & = 
        (1 - \pi_0^*) \left( \frac{\alpha}{P_{nz} - 1 + \alpha} \right)
\end{split}
\end{align} 

\section{Simulations} \label{sec:simulation}

\begin{table}[!t]
	\centering
\begin{tabular}{llccc}
\hline\hline
\multicolumn{1}{l}{}&\multicolumn{1}{c}{N = 30}&\multicolumn{1}{c}{N = 60}&\multicolumn{1}{c}{N = 120}&\multicolumn{1}{c}{N = 240}\tabularnewline
\hline
{\bfseries Design  1}&&&&\tabularnewline
~~FOSR&3.02 (0.16)&1.19 (0.08)&0.64 (0.04)&0.3 (0.02)\tabularnewline
~~FOSR-PM&2.12 (0.16)&1.2 (0.08)&0.71 (0.05)&0.29 (0.02)\tabularnewline
~~FOSR-DP&0.94 (0.05)&0.4 (0.03)&0.16 (0.01)&0.07 (0.01)\tabularnewline
~~FOSR-DPPM&1.12 (0.06)&0.43 (0.03)&0.17 (0.01)&0.06 (0.01)\tabularnewline
\hline
{\bfseries Design  2}&&&&\tabularnewline
~~FOSR&4.61 (0.22)&2.8 (0.15)&1.38 (0.08)&0.73 (0.04)\tabularnewline
~~FOSR-PM&4.1 (0.22)&3.49 (0.2)&1.94 (0.11)&0.99 (0.07)\tabularnewline
~~FOSR-DP&1.49 (0.09)&0.89 (0.06)&0.37 (0.03)&0.16 (0.01)\tabularnewline
~~FOSR-DPPM&2.1 (0.12)&1.13 (0.07)&0.48 (0.03)&0.19 (0.02)\tabularnewline
\hline
{\bfseries Design  3}&&&&\tabularnewline
~~FOSR&2.75 (0.1)&1.18 (0.03)&0.64 (0.02)&0.31 (0.01)\tabularnewline
~~FOSR-PM&2.49 (0.08)&1.53 (0.04)&0.81 (0.04)&0.36 (0.01)\tabularnewline
~~FOSR-DP&1.85 (0.05)&1.18 (0.03)&0.75 (0.03)&0.41 (0.01)\tabularnewline
~~FOSR-DPPM&2.16 (0.06)&1.27 (0.04)&0.78 (0.03)&0.42 (0.02)\tabularnewline
\hline
{\bfseries Design  4}&&&&\tabularnewline
~~FOSR&1.73 (0.07)&0.7 (0.03)&0.31 (0.01)&0.16 (0.01)\tabularnewline
~~FOSR-PM&1.09 (0.03)&0.67 (0.03)&0.29 (0.01)&0.14 (0.01)\tabularnewline
~~FOSR-DP&0.95 (0.03)&0.62 (0.02)&0.3 (0.01)&0.17 (0.01)\tabularnewline
~~FOSR-DPPM&1.16 (0.04)&0.62 (0.02)&0.3 (0.01)&0.17 (0.01)\tabularnewline
\hline
\end{tabular}
 
	\caption{Mean pointwise MSE for 100 datasets for each model. Standard errors (in parentheses) 
		are
		estimated using the bootstrap with 100 repetitions. }
	\label{tab:mse_mean}
\end{table}

\begin{table}
	\centering
	\begin{subtable}[t]{\linewidth}
		\centering
		\caption{RAND Index}
		\vspace{0pt}
		\begin{tabular}{llccc}
			\hline\hline
			\multicolumn{1}{l}{}&\multicolumn{1}{c}{N = 30}&\multicolumn{1}{c}{N = 
				60}&\multicolumn{1}{c}{N = 120}&\multicolumn{1}{c}{N = 240}\tabularnewline
			\hline
			{\bfseries Design  1}&&&&\tabularnewline
			~~FOSR-DP&0.69 (0.01)&0.79 (0.01)&0.92 (0.01)&0.96 (0.01)\tabularnewline
			~~FOSR-DPPM&0.65 (0.01)&0.8 (0.01)&0.92 (0.01)&0.97 (0.01)\tabularnewline
			\hline
			{\bfseries Design  2}&&&&\tabularnewline
			~~FOSR-DP&0.69 (0.01)&0.78 (0.01)&0.87 (0.01)&0.93 (0.01)\tabularnewline
			~~FOSR-DPPM&0.63 (0.01)&0.76 (0.01)&0.84 (0.01)&0.92 (0.01)\tabularnewline
			\hline
		\end{tabular}
	\end{subtable} \\
	\vspace{0.125in}
	\begin{subtable}[t]{\linewidth}
		\centering
		\caption{Adjusted RAND Index}
		\begin{tabular}{llccc}
			\hline\hline
			\multicolumn{1}{l}{}&\multicolumn{1}{c}{N = 30}&\multicolumn{1}{c}{N = 
				60}&\multicolumn{1}{c}{N 
				= 
				120}&\multicolumn{1}{c}{N = 240}\tabularnewline
			\hline
			{\bfseries Design  1}&&&&\tabularnewline
			~~FOSR-DP&0.41 (0.02)&0.59 (0.02)&0.83 (0.02)&0.93 (0.01)\tabularnewline
			~~FOSR-DPPM&0.47 (0.03)&0.64 (0.02)&0.84 (0.02)&0.94 (0.01)\tabularnewline
			\hline
			{\bfseries Design  2}&&&&\tabularnewline
			~~FOSR-DP&0.43 (0.02)&0.57 (0.03)&0.78 (0.02)&0.87 (0.02)\tabularnewline
			~~FOSR-DPPM&0.22 (0.02)&0.45 (0.03)&0.64 (0.02)&0.87 (0.02)\tabularnewline
			\hline
		\end{tabular}
	\end{subtable}
	\caption{Mean RAND and Adjusted RAND index for 100 datasets is
		reported for each model. Standard errors (in parentheses) are
		estimated using the bootstrap with 100 repetitions. } \label{tab:clust_mean}
\end{table}

In this section we evaluate the efficacy of our models using a
simulation study. Our simulations consist of four designs and compare the
performance of the proposed priors with two priors from the literature; the
smoothing prior proposed by \cite{goldsmith2016assessing} and a
modification of the variable selection prior proposed by
\cite{goldsmith2017variable}. For clarity, the
priors and the corresponding names used in our tables are listed below: 
\begin{itemize}
	\item FOSR: \cite{goldsmith2016assessing}
	\begin{align*}
		(\btilde_p, \tilde{\lambda}_{b_p}) & \sim 
			\normal_M(\V{0}, \tilde{\lambda}_{b_p}\Inv \R\Inv)
			\mathcal{G}(a_\lambda, b_\lambda)
	\end{align*}
	\item FOSR-PM: Modification of \cite{goldsmith2017variable}	to include smoothing
	\begin{align*}
		(\btilde_p, \tilde{\lambda}_{b_p}) & \sim 
			\pi_0 \delta_{\V{0}} + (1 - \pi_0) G \text{ with } G = 
			\normal_M(\V{0}, \tilde{\lambda}_{b_p}\Inv \R\Inv) \
			\mathcal{G}(a_\lambda, b_\lambda)
	\end{align*}
	\item FOSR-DP: Clustering prior given in \eqref{eqn:prior_dp}
	\item FOSR-DPPM: Variable selection and clustering prior given in \eqref{eqn:prior_dppm}
\end{itemize}

\subsection{Design} 
All of the methods considered are fit under the assumption of iid errors, even
though our simulations use an exponential covariance matrix with a nugget term.
Using the notation outlined in Equation \eqref{eqn:dense_model}, we set $T =
15$, $P_f = 5$, and $P_c = 15$ for all simulations, and use three Fourier
basis functions to generate the coefficient functions. Additionally, we generate
correlated values for $\X$ with $Cov(x_{i, p}, x_{i, p'}) = 0.75^{|p - p'|}$,
while $w_{ij} \simiid \normal(0, 1)$.  Finally, we simulate each dataset with
$\msigma = \msigma' + \sigma^2 \I$ where $\msigma'_{i, j} = \exp (- 10 * (i -
j)^2)$ and $\sigma^2$ was chosen to set the signal-to-noise ratio in the data to
1. The major variation in our simulation designs is the
cluster membership pattern of $\vbeta$; it is varied as follows: 

\begin{itemize}
	\item \textbf{Design 1}: $\vbeta$ has three clusters with 
		$\c= (\V{1}_7\Tra, \V{2}_4\Tra, \V{3}_4\Tra)\Tra$. 
	The coefficients of cluster 1 are set to zero. 
	\item \textbf{Design 2}: $\vbeta$ has three clusters with the same pattern 
		as Design 1. However, all of the effects are non-zero.  
	\item \textbf{Design 3}: $\vbeta$ has no clusters and none of the coefficients are zero. 
	\item \textbf{Design 4}: Similar to Design 3 but the first seven elements of $\vbeta$ are 
		set to zero. 
	
\end{itemize}

\subsection{Results} 

Each model was estimated using the Gibbs sampler described in Section
\ref{sec:computation}, run for 5000 iterations with 2500 used as burn-in.
Additionally, we used eight b-spline basis functions to estimate each
coefficient function.  We evaluate our methods using two criteria: parameter
estimation using the pointwise mean squared error (MSE) of the functional
effects, and clustering ability using the RAND and adjusted RAND indices. 

Table \ref{tab:mse_mean} shows the pointwise MSE of our proposed methods in
comparison to the other priors considered. It can be seen that when cluster
structure is present in the coefficients, as in Designs 1 and 2, the FOSR-DP and
FOSR-DPPM methods outperform FOSR and FOSR-PM handily. This is to be expected
since our priors are designed to exploit this structure. What is interesting
however, is the out-performance of our methods in the small sample setting for
Designs 3 and 4. The data in Design 3 is simulated to give FOSR an
advantage, however both FOSR-DP and FOSR-DPPM outperform FOSR
when $N = 30$. Since the total number of predictors in the simulations is fixed
at 20, and none are set to zero in Design 3, the high dimensionality of problem
seems to lend itself well to the use of a dimension
reduction method which borrows information from 
nearby covariates.
We see something similar in Design 4, where FOSR-DP outperforms
the FOSR-PM prior. 

We also evaluated the clustering capacity of our methods using the RAND and
Adjusted RAND indices, with results in Table \ref{tab:clust_mean}. As can be
seen, our methods are able to learn cluster structure in the coefficients despite
miss-specification of the covariance matrix and a low signal-to-noise ratio.

\section{Application} \label{sec:application}

\begin{figure} 
	\centering
	\includegraphics[scale=0.4]{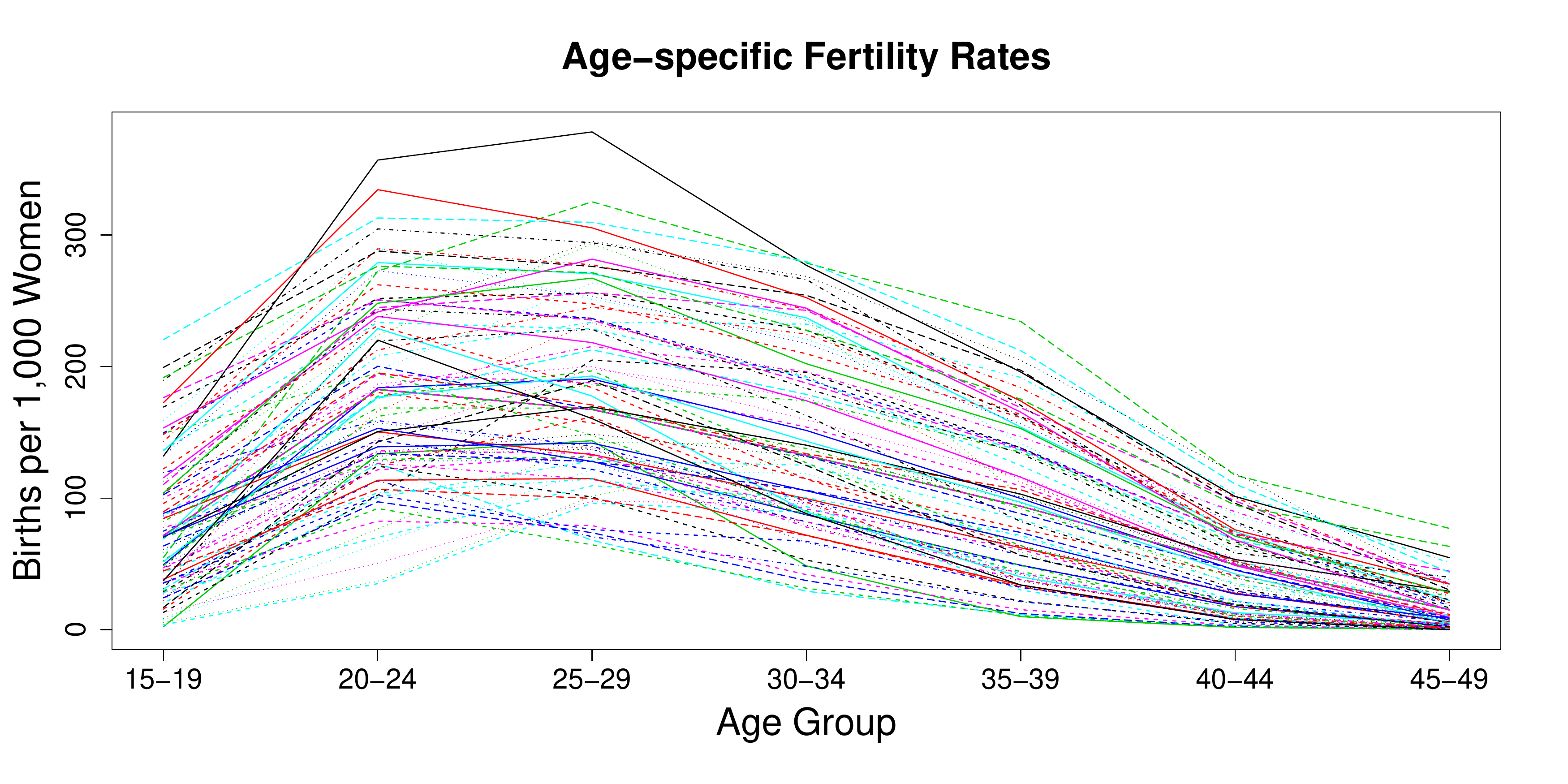}
	\caption{Age-specific fertility rates for 92 countries from the United Nations Gender Information 
		Database} \label{fig:asfr_curves}
\end{figure}

We demonstrate the use of our methods by analyzing age-specific fertility rates
from the United Nations Gender Information (UNGEN) database. We use data from
surveys conducted between 2000-2005, estimates of which are plotted in Figure
\ref{fig:asfr_curves}. Due to the nature of the survey, fertility curves are
only observed at 7 points, where each time point corresponds to an age
group for the participants in question. Each curve seems to be a relatively
smooth function of time, making a function-on-scalar analysis of this data
appropriate. A similar data set is explored by \cite{kowal2018dynamic} but our
analysis considers a larger population of countries and a greater number of
demographic and socioeconomic predictors. 

The response information from the UNGEN database is combined with fifteen
country level covariates available on Gapminder. Each covariate included in the
final model is a complete case average of Gapminder data from 2000-2005. This
allows us to see how average values of demographic and socioeconomic variables
during the years of the UN survey affect the fertility curve. The variables we
consider include some potentially causal factors such as age at first
marriage, something which should increase fertility in the early parts of the
curve, along with seemingly inconsequential variables such as the proportion of
dollar billionaires in a country and the amount of alcohol its inhabitants
consume. Unfortunately, many countries had no covariate information
for the years of the survey and were dropped from the analysis for
simplicity.  The final data set consists of 15 covariates for 92 countries
around the world. 

\begin{figure}
	\centering
	\hspace{-0.75in}
	\includegraphics[scale = 0.45]{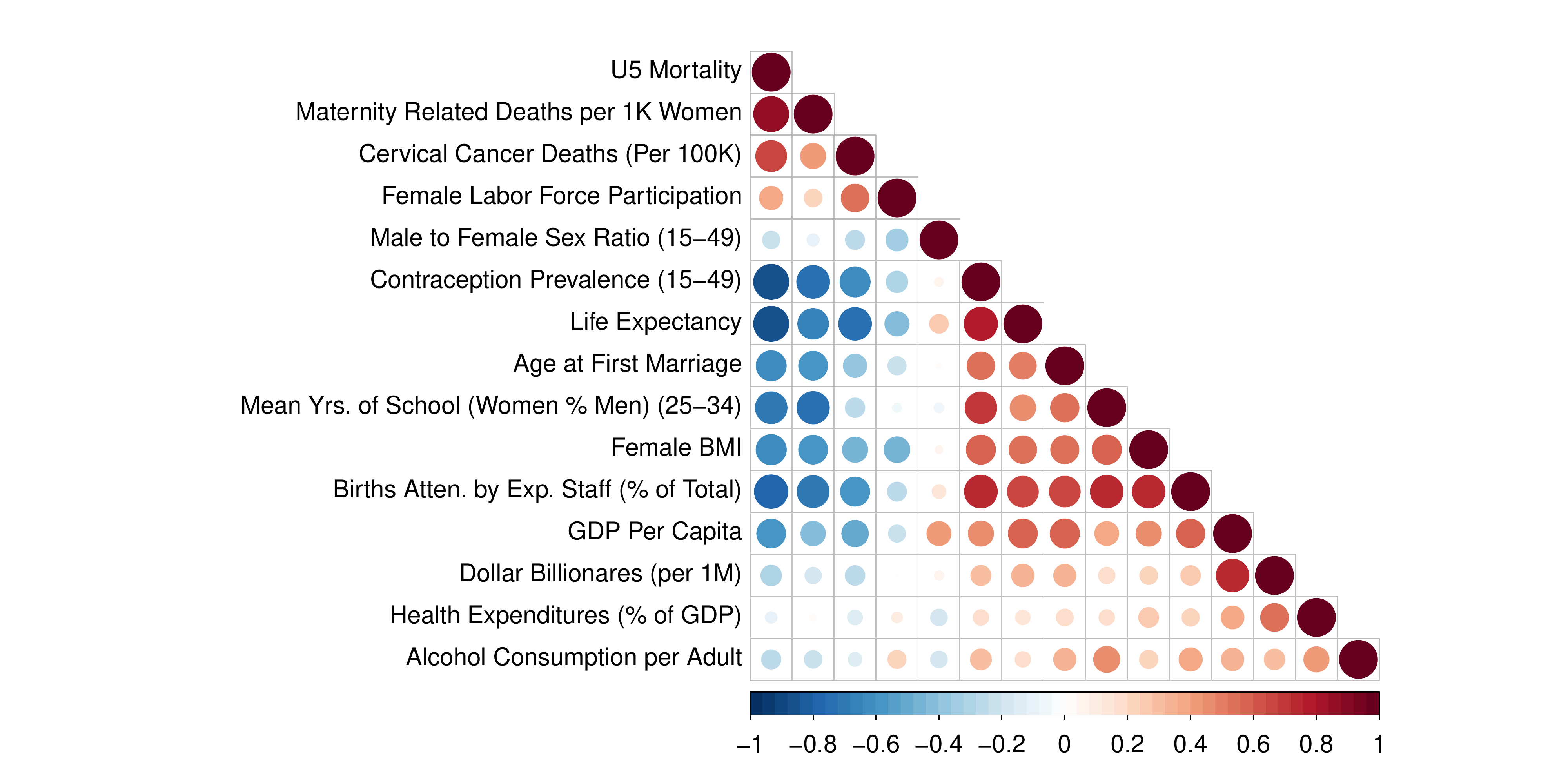}
	\caption{Plot of the correlation matrix of predictors used
		to fit the FOSR-DP and FOSR-DPPM models} \label{fig:corr_plot}
\end{figure}

Figure \ref{fig:corr_plot} shows many highly correlated predictors among
consideration. For example, Under-5 Mortality is highly positively correlated
with Maternal Deaths (correlation coefficient of 0.85), but is highly negatively
correlated with Contraception Prevalence (correlation of -0.85). Contraception
Prevalence on the other hand is positively correlated with many socioeconomic 
factors such
as life expectancy and the number of births attended by trained birth staff.

\begin{figure} 
	\centering
	\includegraphics[scale=0.3]{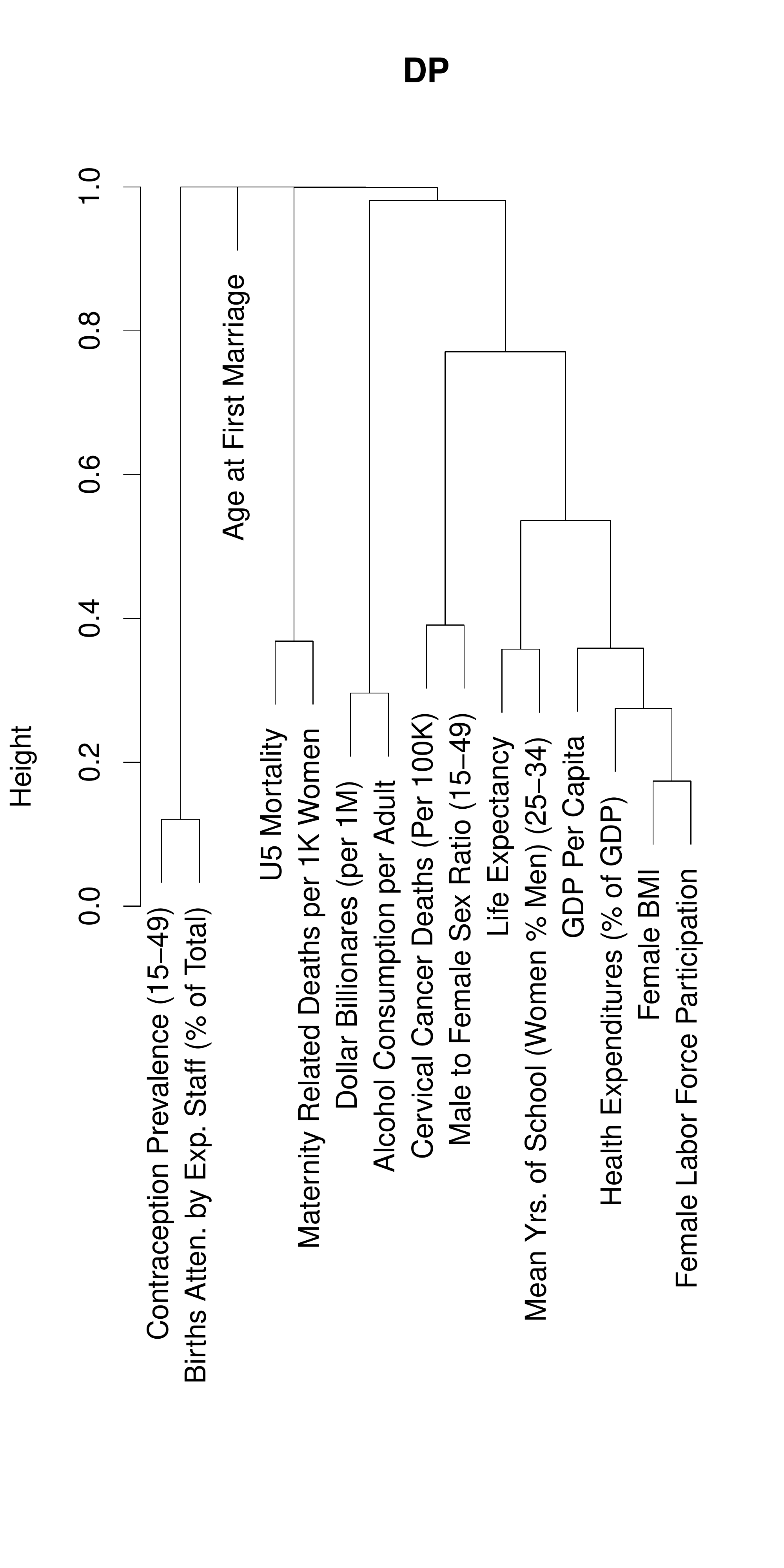}
	\includegraphics[scale = 0.3]{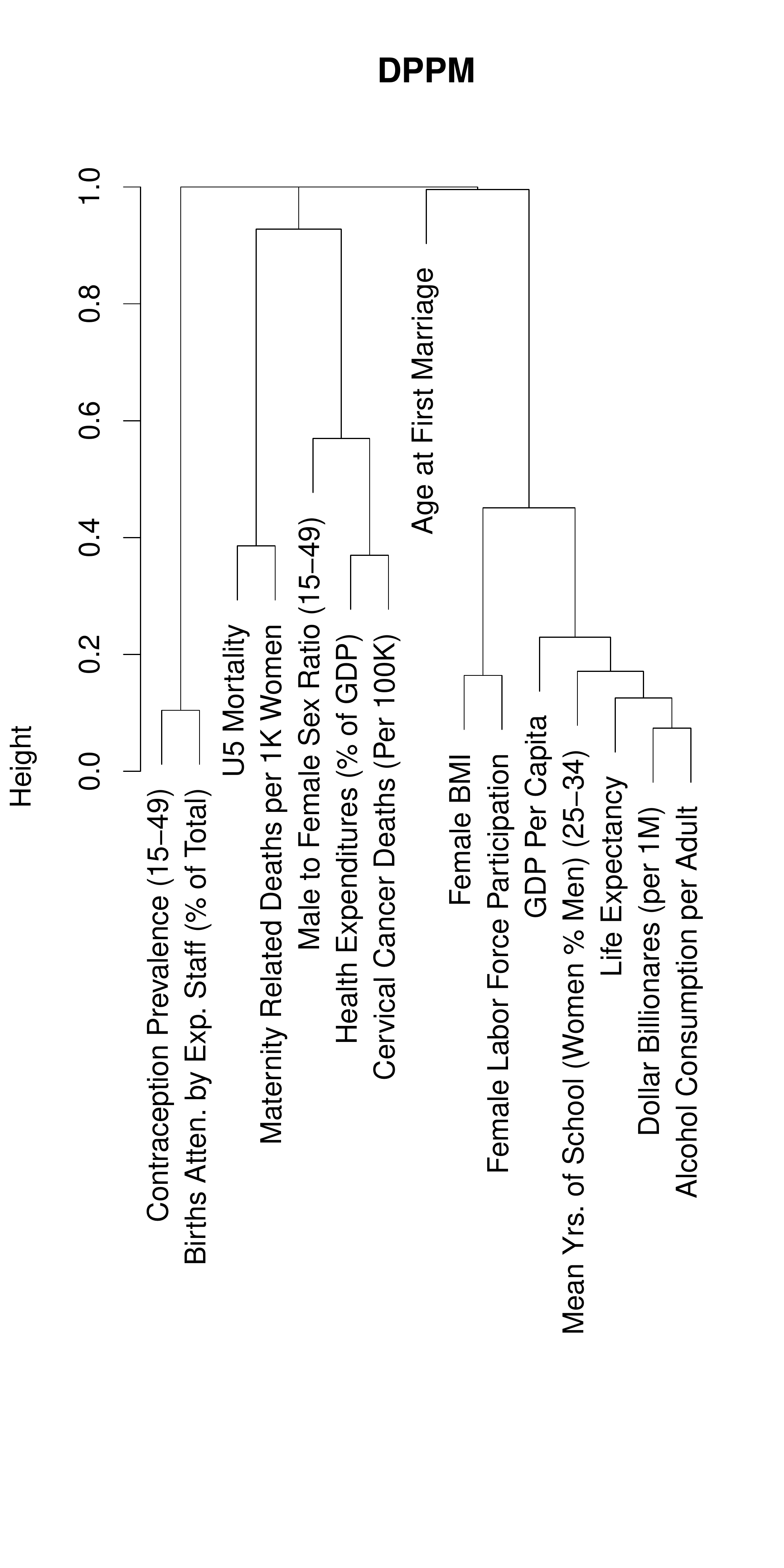}
	\vspace{-0.5in}
	\caption{Dendograms from the DP and DPPM clust models} \label{fig:clust_dend}
\end{figure}

We fit the FOSR-DP and FOSR-DPPM models on the data set. Each method was fit
using 10,000 iterations with 5,000 used for burn-in. Figure \ref{fig:clust_dend}
show the cluster dendogram for the analyses. Viewing both dendograms we see that
covariates with high positive correlation tend to be clustered together.
Additionally, the structure of the tree seems similar across the two models,
with small differences in the order that they are joined in a particular node.
Finally, the variable that we thought a-priori should most clearly impact the
age-specific fertility curve, age at first marriage, is a standalone covariate
in both dendograms. Table \ref{tab:perc_zero} shows the percent of iterations
after burn-in that each covariate was set to zero by the FOSR-DPPM model; of the
fifteen covariates considered, only six are estimated to have non-zero effects
if a 5\% cutoff is used. 

\begin{table}
\begin{center}
\begin{tabular}{lr}
\hline\hline
\multicolumn{1}{l}{}&\multicolumn{1}{c}{Percent Zero}\tabularnewline
\hline
Contraception Prevalence (15-49)&$0.000$\tabularnewline
Maternity Related Deaths per 1K Women&$0.000$\tabularnewline
Births Atten. by Exp. Staff (\% of Total)&$0.000$\tabularnewline
Age at First Marriage&$0.006$\tabularnewline
Male to Female Sex Ratio (15-49)&$0.007$\tabularnewline
U5 Mortality&$0.012$\tabularnewline
Cervical Cancer Deaths (Per 100K)&$0.148$\tabularnewline
Health Expenditures (\% of GDP)&$0.291$\tabularnewline
Female BMI&$0.556$\tabularnewline
Female Labor Force Participation&$0.579$\tabularnewline
GDP Per Capita&$0.814$\tabularnewline
Mean Yrs. of School (Women \% Men) (25-34)&$0.861$\tabularnewline
Life Expectancy&$0.901$\tabularnewline
Dollar Billionares (per 1M)&$0.948$\tabularnewline
Alcohol Consumption per Adult&$0.959$\tabularnewline
\hline
\end{tabular}\end{center}

	\caption{Shows the proportion of iterations that the variables listed were set to zero by the 
		FOSR-DPPM model} \label{tab:perc_zero}
\end{table}

\begin{figure}
	\centering
	\includegraphics[scale=0.25]{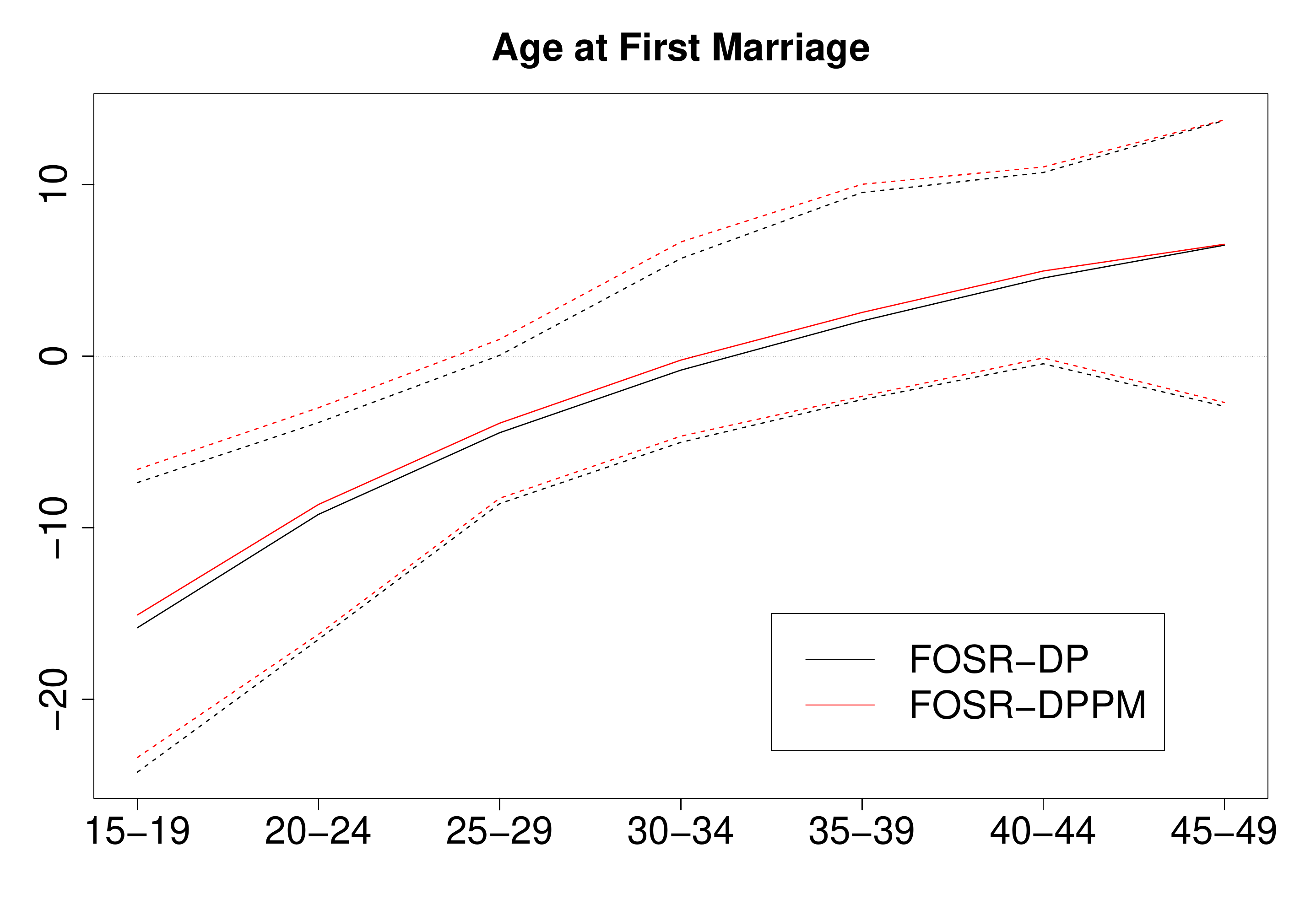}
	\includegraphics[scale=0.25]{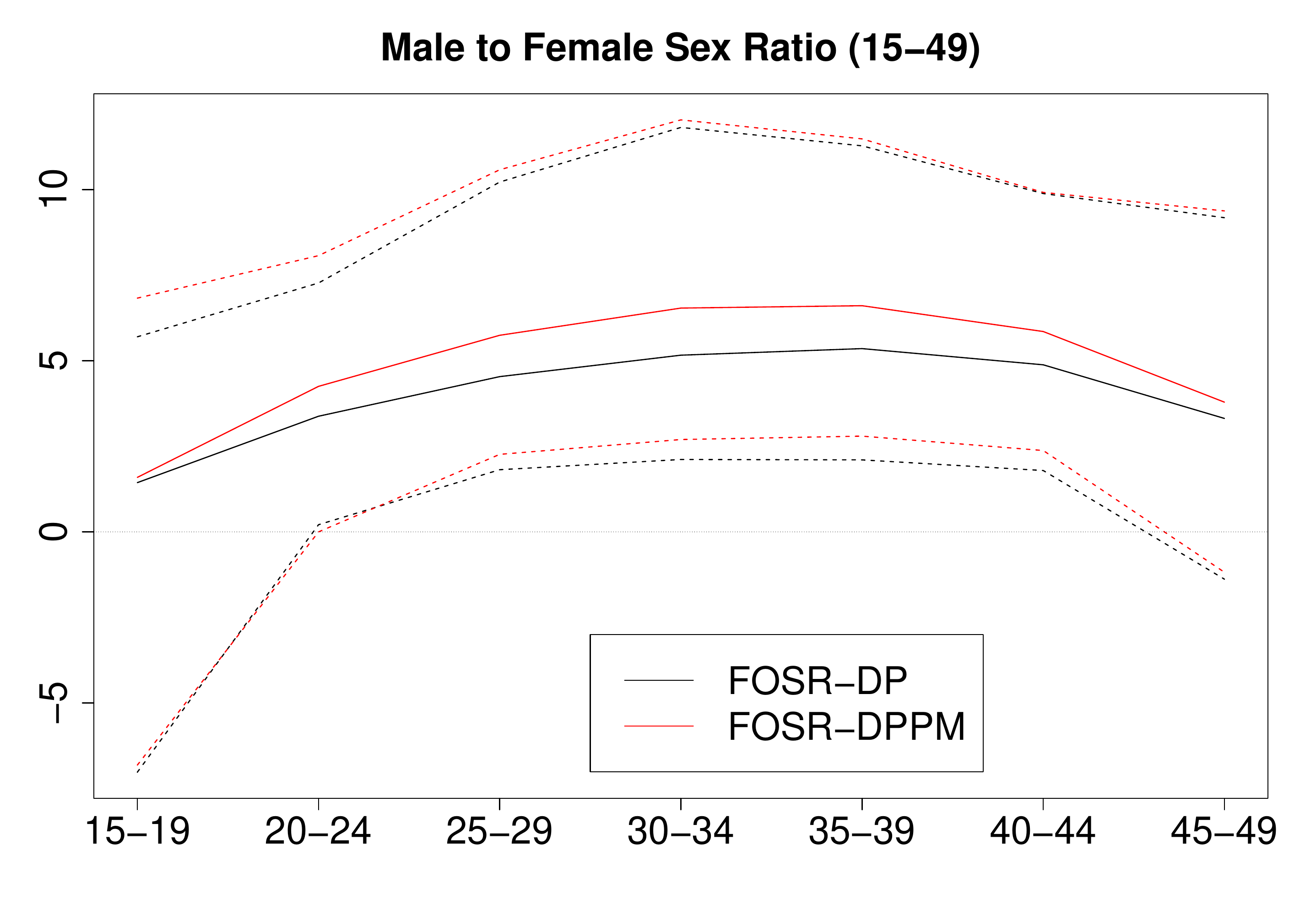}	
	\includegraphics[scale=0.25]{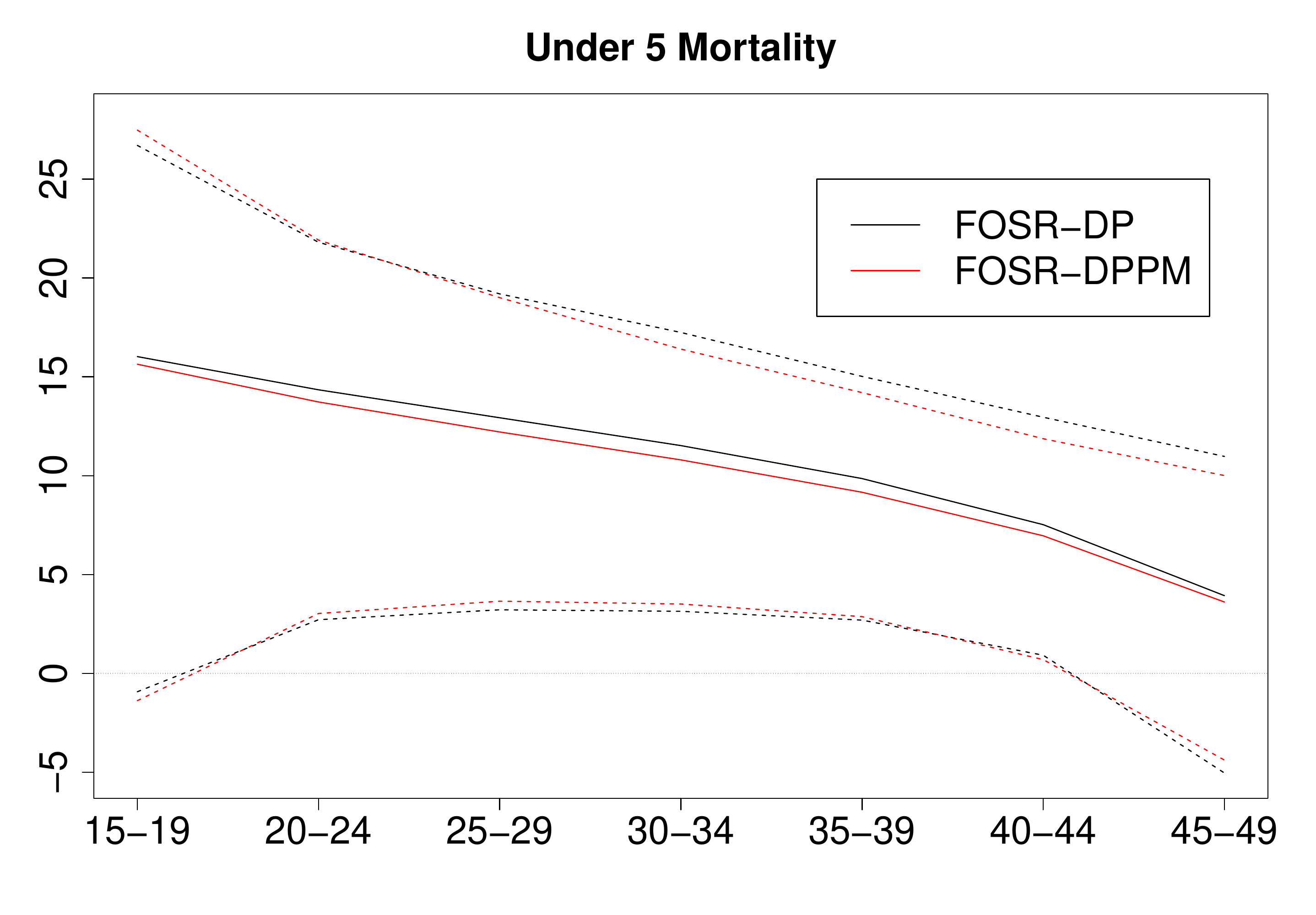}
	\includegraphics[scale=0.25]{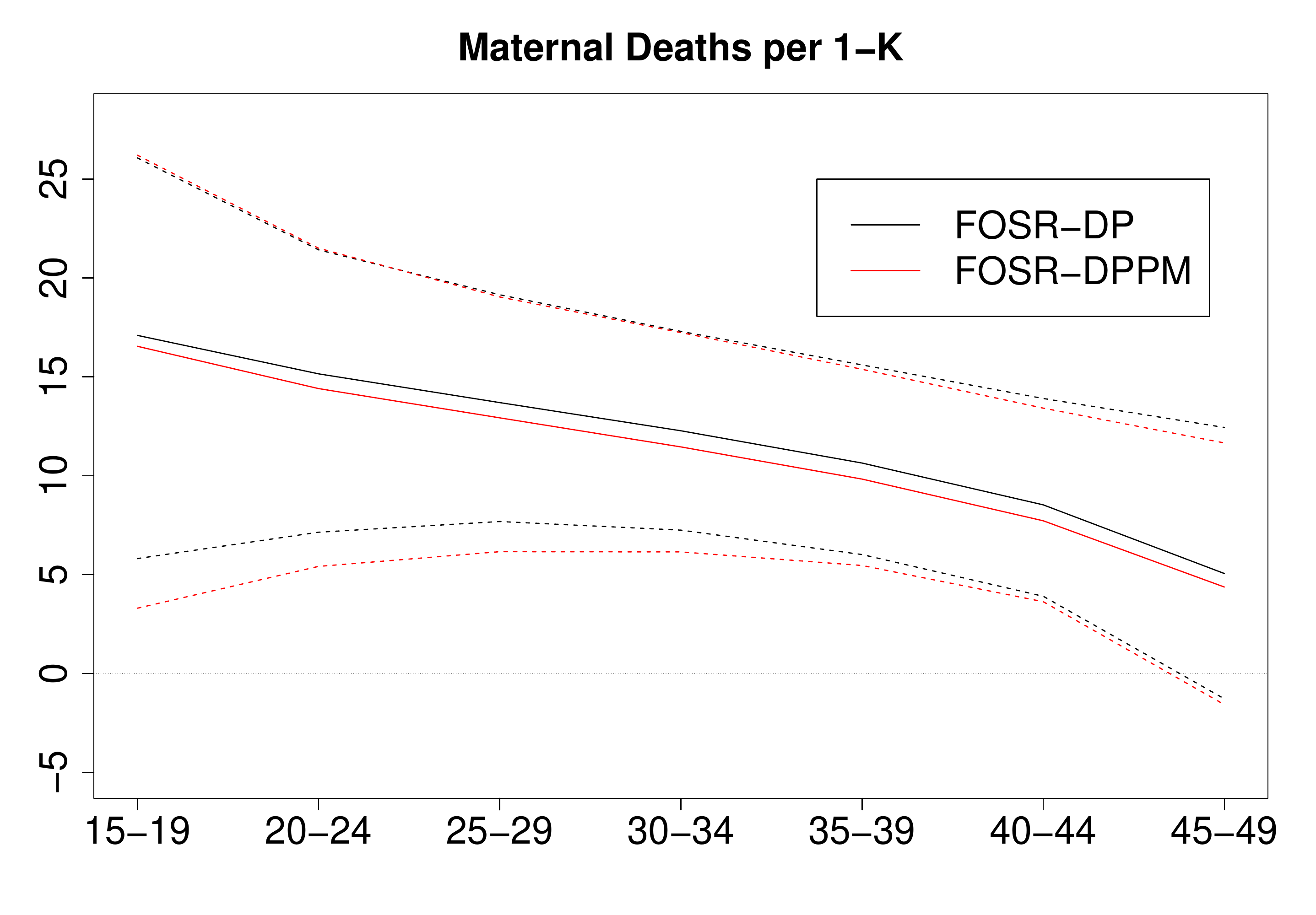}
	\includegraphics[scale=0.25]{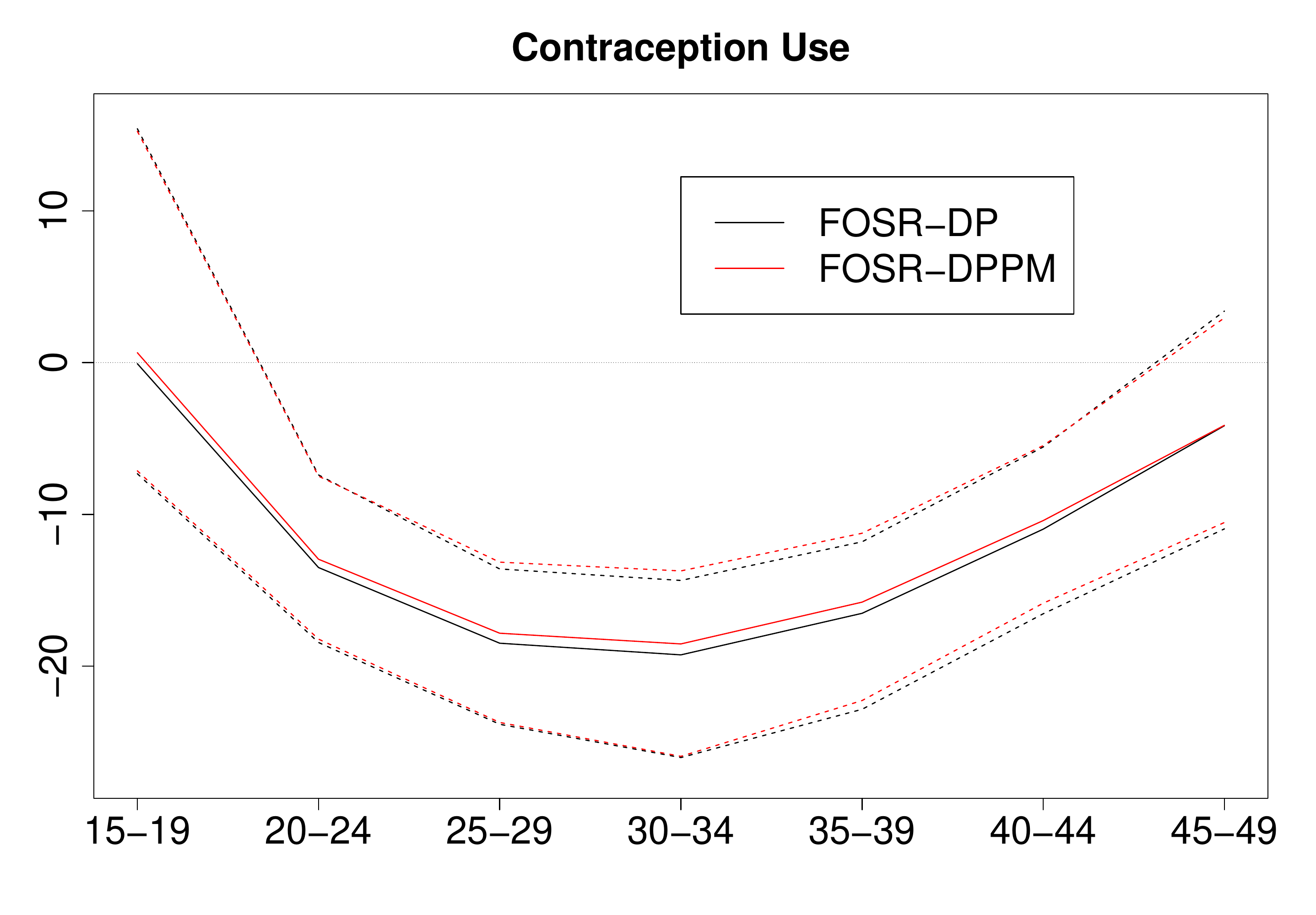}
	\includegraphics[scale=0.25]{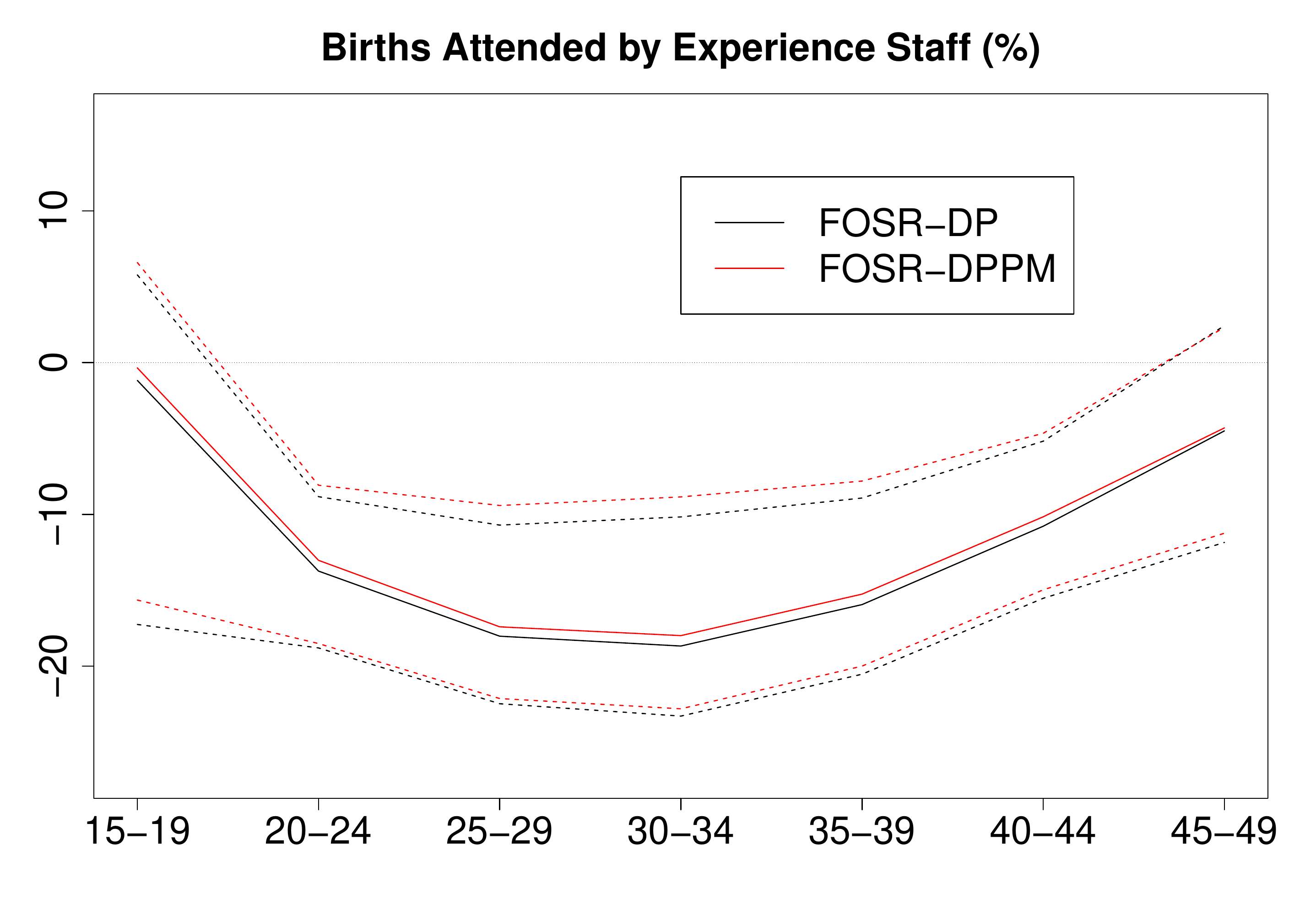}
	\caption{Plots showing the estimated coefficient  curves from the FOSR-DP and FOSR-DPPM 
	models. The solid lines are the posterior means of 5000 samples from the Gibbs sampler, while 
	the dotted lines are the 2.5 and 97.5 percentile respectively} \label{fig:func_effects}
\end{figure}

Finally, the estimated effects, on a standardized scale, of the non-zero coefficients are shown in Figure
\ref{fig:func_effects}.  From this we can see that the age at first marriage has
a time varying effect; a decrease of one standard deviation of age increases the
number of births in the 15-19 and 20-24 age groups, but the effect disappears as
age increases. Additionally, U5 mortality and Maternal Deaths per 1-K are
highly correlated variables with clustered coefficient curves. Their effect is
positive throughout, but decreases as age increases. Finally, an interesting
finding in our estimates is that contraception prevalence primarily impacts fertility
in the middle of the curve; it is not impactful in the 15-19 age group. 

\section{Conclusion}

In this paper we extend methodology for addressing multicollinearity in the
predictors to function-on-scalar regression. We contribute to the literature by
proposing a prior which simultaneously selects, clusters, and smooths the
coefficient functions using a Bayesian approach, while the current Bayesian
literature only focuses independently on selection or smoothing. Our model allows
coefficient estimates to remain flexible while controlling overfitting, 
and performs dimension reduction by summing
columns in the covariate matrix which have the same effect on the response. We
also develop a Gibbs sampler which converges to the posterior distribution
quickly in practice by combining multiple sampling algorithms developed in the
clustering literature.  

Our work can be extended in many ways. First, the Gibbs sampler proposed in
Section \ref{sec:computation} is not scalable to problems with large $N$ or $P$,
because collapsing the sampler over the basis function coefficients forces us to
solve many linear systems in each iteration. Faster methods utilizing other
algorithms for Dirichlet Process priors, with a focus on Variational Bayes
approaches, have the potential to make our methodology applicable to larger data
sets. Second, our methods assume that the coefficient curves are globally
clustered, which can be a strong assumption in practice. Finally, if point
estimation is the only goal, methods such as OSCAR, the Clustered Lasso, and
PACS can be extended to the function-on-scalar regression framework, possibly
providing another computationally efficient alternative to our approach.

\bibliography{references} 
\bibliographystyle{plainnat}

\end{document}